\documentclass[onecolumn]{aa} 
\usepackage{graphicx}
\usepackage{natbib}
\usepackage{deluxetable}
\usepackage{txfonts}
\begin{document}
   \title{Structural and compositional properties of brown dwarf disks: the case of 2MASS~J04442713+2512164}

   \author{H. Bouy\inst{1,2} \thanks{Marie Curie Outgoing International Fellow MOIF-CT-2005-8389}
          \and N. Hu\'elamo\inst{3}
          \and C. Pinte\inst{4}
          \and J. Olofsson\inst{5}
	  \and D. Barrado y Navascu\'es\inst{3}
	  \and E.~L. Mart\'\i n\inst{2}
          \and E. Pantin\inst{6}
          \and J.-L. Monin\inst{5}
	  \and G. Basri\inst{1}
          \and J.-C. Augereau\inst{5}
          \and F. M\'enard\inst{5}
          \and G. Duvert\inst{5}
          \and G. Duch\^ene\inst{1}
	  \and F. Marchis\inst{1}
	  \and A. Bayo\inst{3}
	  \and S. Bottinelli\inst{7}
          \and B. Lefort\inst{8}
	  \and S. Guieu\inst{9}
          }

   \offprints{H. Bouy}

   \institute{Astronomy Department, University of California, Berkeley, CA 94720, USA\\
     \email{hbouy@astro.berkeley.edu}
     \and
     Instituto de Astrof\'\i sica de Canarias, C/ V\'\i a L\'actea s/n, E-38200 - La Laguna, Tenerife, Spain
     \and
     Laboratorio de Astrof\'\i sca Espacial y F\'\i sica Fundamental (LAEFF-INTA), Apdo 50727, 28080 Madrid, Spain
     \and
     Astrophysics Group, University of Exeter, Stocker Road, Exeter, EX4 4QL, United Kingdom
     \and
     Laboratoire d'Astrophysique de Grenoble, CNRS, Universit\'e Joseph Fourier,, UMR 5571, BP 53, F-38041 Grenoble, Cedex 9, France     
     \and
     Services d'Astrophysique Saclay CEA F-91191 Gif-sur-Yvette France
     \and Leiden Observatory, P.O. Box 9513, 2300 RA Leiden, Netherlands
     \and Grantecan S.A., C/ V\'\i a L\'actea s/n, E-38200 - La Laguna, Tenerife, Spain
     \and Spitzer Science Center, California Institute of Technology, Mail Code 314-6, 1200 East California Boulevard, Pasadena, CA 91125, USA
   }

   \date{Received September 15, 1996; accepted March 16, 1997}

 
  \abstract
   {}
   {In order to improve our understanding of substellar formation, we have performed a compositional and structural study of a brown dwarf disk.}
   {We present the result of photometric, spectroscopic and imaging observations of 2MASS~J04442713+2512164, a young brown dwarf (M7.25) member of the Taurus association. Our dataset, combined with results from the literature, provides a complete coverage of the spectral energy distribution from the optical to the millimeter including the first photometric measurement of a brown dwarf disk at 3.7~mm, and allows us to perform a detailed analysis of the disk properties.}
   {The target was known to have a disk. High resolution optical spectroscopy shows that it is intensely accreting, and powers a jet and an outflow. The disk structure is similar to that observed for more massive TTauri stars. Spectral decomposition models of Spitzer/IRS spectra suggest that the mid-infrared emission from the optically thin disk layers is dominated by grains with intermediate sizes (1.5~$\mu$m). Crystalline silicates are significantly more abondant in the outer part and/or deeper layers of the disk, implying very efficient mixing and/or additional annealing processes. Sub-millimeter and millimeter data indicate that most of the disk mass is in large grains ($>$1~mm).}
   {}

   \keywords{Physical data and processes: Radiative transfer, Stars: circumstellar matter, Stars: formation, Stars: individual: 2MASS~J04442713+2512164, Stars: brown dwarfs               }

   \maketitle
%

\section{Introduction}
In the last decade, several studies have confirmed the presence of circumstellar disks around very low-mass stars and brown dwarfs \citep[see e.g][ and references therein for a recent review]{2007astro.ph..2286A}.  These objects undergo an accretion phase, like the more massive TTauri stars (hereafter TTauri), reinforcing the idea that they mostly form in the same way as stars, i.e. via fragmentation and collapse of a molecular cloud. 

Observational evidence for accretion and mass-loss at and below the sub-stellar limit have been obtained for a growing number of sources in various star forming regions \citep[e.g ][]{2001A&A...380..264F,2005A&A...440.1119F,2004ApJ...604..284B,2005ApJ...626..498M,2007ApJ...659L..45W}. \citet{2004A&A...424..603N} have studied the accretion properties of a combined sample of  20 sub-stellar objects for which the mass-accretion rate is known.  These results have shown that the trend of lower accretion rates for lower-mass objects continues in the sub-stellar regime. This trend has been confirmed in subsequent work on larger samples of very low mass and sub-stellar objects by e.g \citet{2005ApJ...626..498M} and \citet{2005ApJ...625..906M}.

Taking advantage of the unprecedented sensitivity of the Spitzer Space Telescope, it is now possible to investigate the composition of very low mass star and brown dwarf disks as previously done for more massive TTauri and Herbig AeBe objects \citep{2005Sci...310..834A,2007ApJ...661..361M}. While the overall disk properties seem to be similar on each side of the sub-stellar limit, significant discrepancies are observed when looking at individual targets.

In this paper, we present new observations of a young brown dwarf member of the Taurus association. The mid-IR and millimeter excesses reported respectively by \citet{2004AJ....127.3553K,2007A&A...465..855G,2006ApJ...645.1498S} indicate that the object harbors a disk. We present and analyze archival sub-millimeter and new mid-IR photometry and spectroscopy data and perform a detailed analysis of the structure of the disk via spectral energy distribution (SED) modelling, and of its composition and mineralogy via the modelling of the mid-IR spectrum.

\section{Observations}
The target, 2MASS~J04442713+2512164, has been initially detected with IRAS by \citet{2004AJ....127.3553K} who identified it as a young stellar object. It was subsequently classified as a M7.25$\pm$0.25 brown dwarf member of the Taurus association by \citet{2004ApJ...617.1216L}. In their optical spectrum, they report strong forbidden emission lines indicative of the presence of a jet. The strong H$\alpha$ emission present in their spectrum also classifies the object as an accretor. Now, we have obtained and compiled a comprehensive dataset covering the electromagnetic spectrum from 0.43~$\mu$m to 3.7~mm in order to understand the properties of this object.

\subsection{High resolution optical spectroscopy with Keck/HIRES}
We have obtained a high resolution optical spectrum using HIRESr \citep{1994SPIE.2198..362V} on Keck I on 2006 October 13th. We used a setting that covers the wavelength region from 5600 to 10\,000\AA, and a slit width of 1.15\arcsec, yielding a resolution of R=31\,000. We obtained a single exposure of 40~min. The spectrum was processed and extracted using standard procedures with custom routines under IDL (Interactive Data Language). Fig. \ref{lines} shows some of the major spectral features observed in the spectrum.

\subsection{Low resolution optical spectroscopy with CAHA/TWIN}
We have obtained low resolution (R=3000, 0.57$\le\lambda\le$0.99~$\mu$m) optical spectra using TWIN at the Calar Alto Observatory on 2006 November 20 and 22. We acquired 7 exposures of 10~min each. A photometric standard was observed after the target, but the ambient conditions were relatively poor and variable, and the photometric calibration not very accurate. The data were processed using standard procedure within IRAF. Fig. \ref{twin} shows the average of the seven individual spectra.

\subsection{High angular resolution near-IR images with VLT/NACO}
We used the adaptive optics (AO) NACO instrument \citep{2002Msngr.107....1B} on the VLT to image the target at high angular resolution. On 2006 December 23 and 24, we obtained images in the H, Ks, NB1.64, NB2.12, and L' filters using the N90C10 dichroic and the S13 camera providing a plate-scale of 0\farcs013 and a total field of view of $\approx$13\arcsec$\times$13\arcsec. Exposure times were 10~min for each of the first four filters, and 1h20~min for the L' band. All these observations have been made in jitter mode. The ambient conditions were good, with a seeing $\sigma\le$0\farcs8 and a coherence time between 10$<\tau_{0}<$60~ms, according to the ESO Ambient Condition Database\footnote{measured in the visible and at the zenith}. The data have been processed using the recommended Eclipse software package \citep{1997Msngr..87...19D}. A point spread function (PSF) reference star (2MASS J04441794+2524513, M4III, Ks=6.0~mag) was observed immediately before or after the target, to be used for PSF subtraction. The Strehl ratio in the Ks band reached 16\%.

\subsection{VLT/FORS2 imaging}
\begin{figure*}
   \centering
   \includegraphics[width=0.6\textwidth]{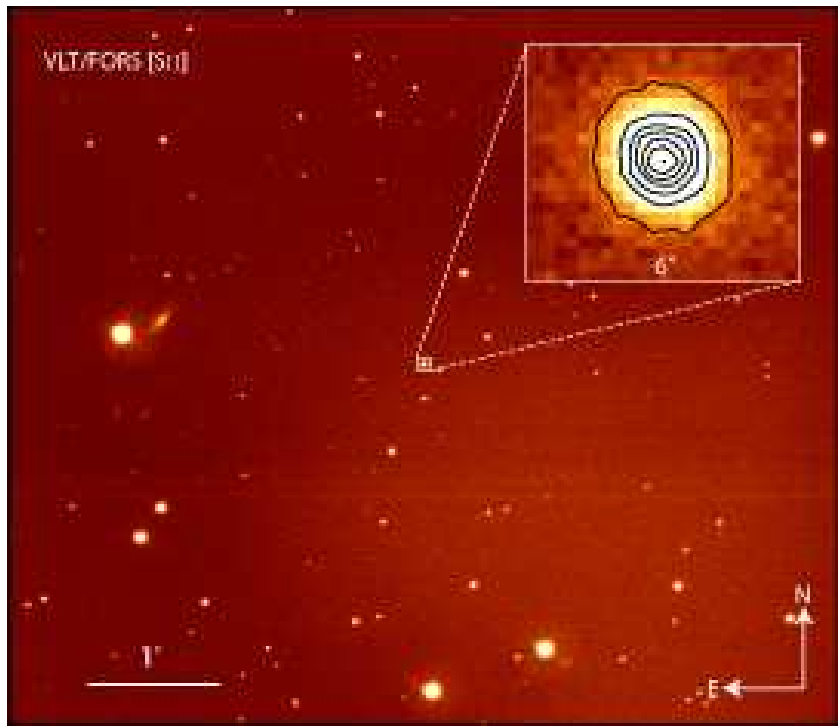}
      \caption{VLT/FORS2 image obtained in the \ion{S}{ii} filter. The target, indicated by a square, is unresolved in this 2~min exposure. The zoom around the target with contour plots shows that there is no clear evidence of a jet. The average seeing in this image is 1\arcsec. The scale and orientation are indicated.}
         \label{fors2}
   \end{figure*}

In order to search for elongations due to the presence of a jet, we have obtained VLT/FORS2 images in the \ion{S}{ii} filter \citep{1998Msngr..94....1A}, where the contrast between such extended features and the star is more favorable. The observations took place on 2007 February 21. We used the standard resolution 2$\times$2 binning mode leading to a image scale of 0\farcs25/pixel resulting in a total field of view of 6.8\arcmin$\times$6.8\arcmin.The pipeline processed image is shown in Fig. \ref{fors2}. The target is unresolved in the 1\arcsec seeing image. No continuum images were taken and no subtraction could be performed.

\subsection{Mid-IR photometry with VLT/VISIR}
We used the mid-infrared imager VISIR \citep{2004Msngr.117...12L} at the VLT to obtain images in the [\ion{Ar}{iii}], PAH1, SIV, PAH2, [\ion{Ne}{ii}], Q1 and Q2 filters. The service mode observations took place between 2006 December 26 and 2007 March 1. Table \ref{visir_table} gives an overview of these filters properties.  The data were processed with a customary pipeline based on IDL  routines. The pipeline includes bad pixel correction and the shift and add of the individual frames, each of them corresponding to a chopping position \citep{2005Msngr.119...25}. The target was detected only in the three shortest wavelength (PAH1, \ion{Ar}{iii} and \ion{S}{iv}) filters. The corresponding photometry is given in Table \ref{photom}. The source is not resolved in any of the images.

\subsection{Millimeter and Sub-millimeter observations}
We searched the JCMT public archive to look for for sub-millimeter (sub-mm) data. 2MASS~J04442713+2512164 has been observed with SCUBA on 2004 December 11th at 450 and 850~$\mu$m (Program ID: M04BH26A2). We retrieved the data, processed them and performed the photometry of the source using standard procedures with the recommended Starlink ORAC software. The pipeline and its recipes are described in \citet{SCUBA_ORACDR}. Following the recommendations in the case of weak sources, all the bolometers were used and median combined to construct the background. Uncertainties were derived using the ``samples'' method. For each integration, the array moves in a small 3\arcsec$\times$3\arcsec jiggle grid, in order to account for a $\approx$1\arcsec offset in the long- and short-wave array centers. The "samples" method views each integration as supplying 9 independent data points and retains all 9$\times$N data points in its final statistics, where N is the number of integrations (N=160=4 scans$\times$40 integrations in our case). The results are given in Table \ref{photom}.

2MASS~J04442713+2512164 was observed on October 12$^{\rm th}$, 2007 in time-filler mode at the IRAM/Plateau de Bure interferometer (Program ID: r01a, P.I. G. Duvert). The array consisted of 4 antennas in an hybrid compact configuration resulting in a beam-size of 9.0$\times$5\arcsec oriented North-South, along P.A=0\degr. The receivers were tunes in SSB (LSB) mode at the central frequency of 86.24GHz ($\lambda$3.48~mm, corresponding to the J=2-1 rotational transition of SiO in its first excited vibrational state), with a 1~GHz bandwidth in the two polarizations. Observing conditions were excellent. The flux and error bars reported in Table \ref{photom} come from a point-source fit in the visibilities at the position of the target. The quality of the observations and the reliability of the IRAM pipeline ensure that the point-source fitting measurement does not suffer from any bias. Four quasars with well known fluxes were observed for the absolute calibration (0507+179, J0418+380, 0528+134, 3C84). The target is detected, making these observations the longest wavelength detection obtained for a sub-stellar object. 

\subsection{Mid-IR spectra (5.8--38~$\mu$m) with Spitzer/IRS}
2MASSJ04442713+2512164 has been observed with the IRS spectrograph \citep{2004SPIE.5487...62H} on-board the Spitzer Space Telescope in staring mode at both short and long wavelength in the course of program 2 (P.I. Houck) on 2005 March 19th. We retrieved the public data and processed them using the \emph{c2d} package which is fully described in \citet[][]{Spizter_IRS_Reduction}. Two different extraction methods are available : full aperture extraction or optimal PSF extraction. The second method is less sensitive to bad pixels or bad data samples than the first method providing spectra with less spikes. In this paper we use the spectrum and corresponding uncertainties obtained with the PSF extraction method. The \emph{c2d} pipeline furthermore corrects for possible pointing errors which can lead to important offsets between different modules. The spectrum ranging from 5 to 38~$\mu$m is in good agreement with the overlapping VISIR photometry within the uncertainties. The spectrum is shown in Fig. \ref{SED}.

\subsection{Previously published measurements}
\citet{2007A&A...465..855G} presented IRAC and MIPS photometry of 2MASSJ04442713+2512164. It was the only brown dwarf of their sample with a detection at 70~$\mu$m. \citet{2006ApJ...645.1498S} presented 1.3~mm observations of 2MASSJ04442713+2512164, and derived a disk mass significantly larger than any other disk mass in their sample. The corresponding photometric measurements are quoted in Table \ref{photom}.

\section{Stellar parameters \label{star}}
Table \ref{table_properties} gives a summary of the stellar parameters of 2MASS~J04442713+2512164.
\vspace{0.5cm}

\begin{figure*}
   \centering
   \includegraphics[width=0.95\textwidth]{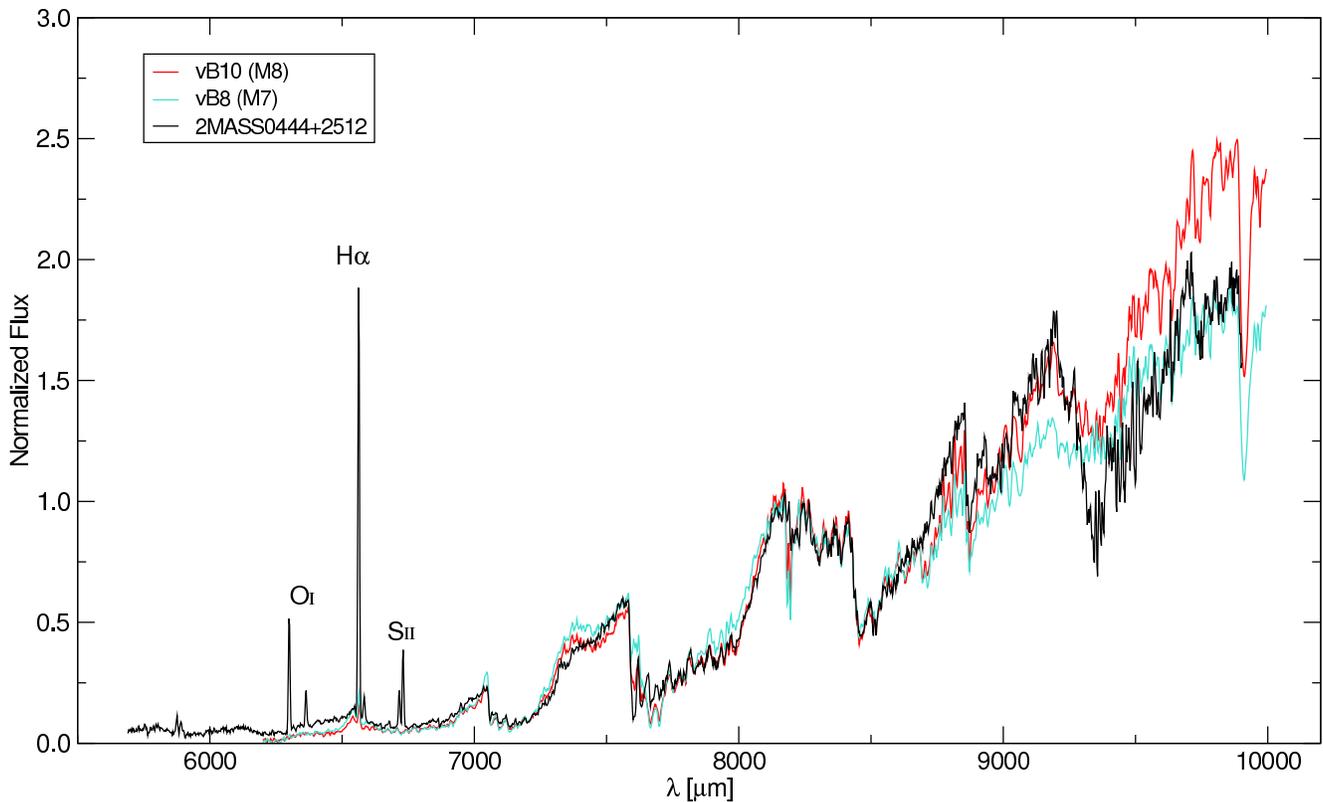}
      \caption{Low resolution CAHA/TWIN optical spectrum of 2MASS~J04442713+2512164 (black), compared to vB8 (M7, light blue) and vB10 (M8, red) field dwarfs observed with the ALFOSC spectrograph on the Northern Optical Telescope at a similar resolution. 2MASS~J04442713+2512164 must have a spectral type intermediate between these two references. \label{twin}  }
   \end{figure*}

\emph{Spectral type. --} Using the low resolution optical spectra obtained with TWIN, and the classification scheme described in \citet{1999AJ....118.2466M}, we derive a PC3 index of 1.64, corresponding to a spectral class of M7.1. Fig. \ref{twin} shows indeed that the spectrum is very similar to that of the field ultra-cool dwarf vB8, which has been classified as an dM7 \citep{1999AJ....118.2466M}. The differences between vB8 and 2MASS~J04442713+2512164 spectra can be ascribed mainly to gravity effects. We thus attribute a final spectral type of M7.0$\pm$0.5. These results are in good agreement with the previous estimates of \citet{2004ApJ...617.1216L}, who reported a spectral type of M7.25$\pm$0.25 and a T$_{\mathrm{eff}}$=2838~K.

\vspace{0.5cm}

\noindent \emph{Rotation velocity. --} We derive the rotation velocity ($v\sin{i}$) of the target by cross-correlating the HIRES spectrum with a grid of ``spun-up'' templates of a slowly rotating standard of similar spectral type obtained the same night \citep[vB~10][]{2006AJ....132..663B}. The cross-correlation was restricted to orders free of strong stellar emission lines, strong telluric features, or gravity-sensitive features. The best match is obtained for $v\sin{i}=$12$\pm$2~km s$^{-1}$.

\vspace{0.5cm}

\noindent \emph{Veiling and extinction. --} From our optical spectrum we derive an extinction A$_{\rm V}$=0, consistent with the previous estimates A$_{\rm J}$=0 of \citet{2004ApJ...617.1216L}. We also estimate the optical continuum veiling $r=\mathcal{F}_{6750\AA}/\mathcal{F}_{6750\AA, cont}$, due to excess emission from the accretion shock, by comparing the strength of the $\lambda$6750~\AA\, TiO molecular band in our high-resolution spectrum to that of a template of similar spectral type and luminosity type (vB10). The best match is obtained for a veiling of 0.4$\pm$0.1.

\vspace{0.5cm}

\begin{figure*}
   \centering
   \includegraphics[width=0.6\textwidth, angle=270]{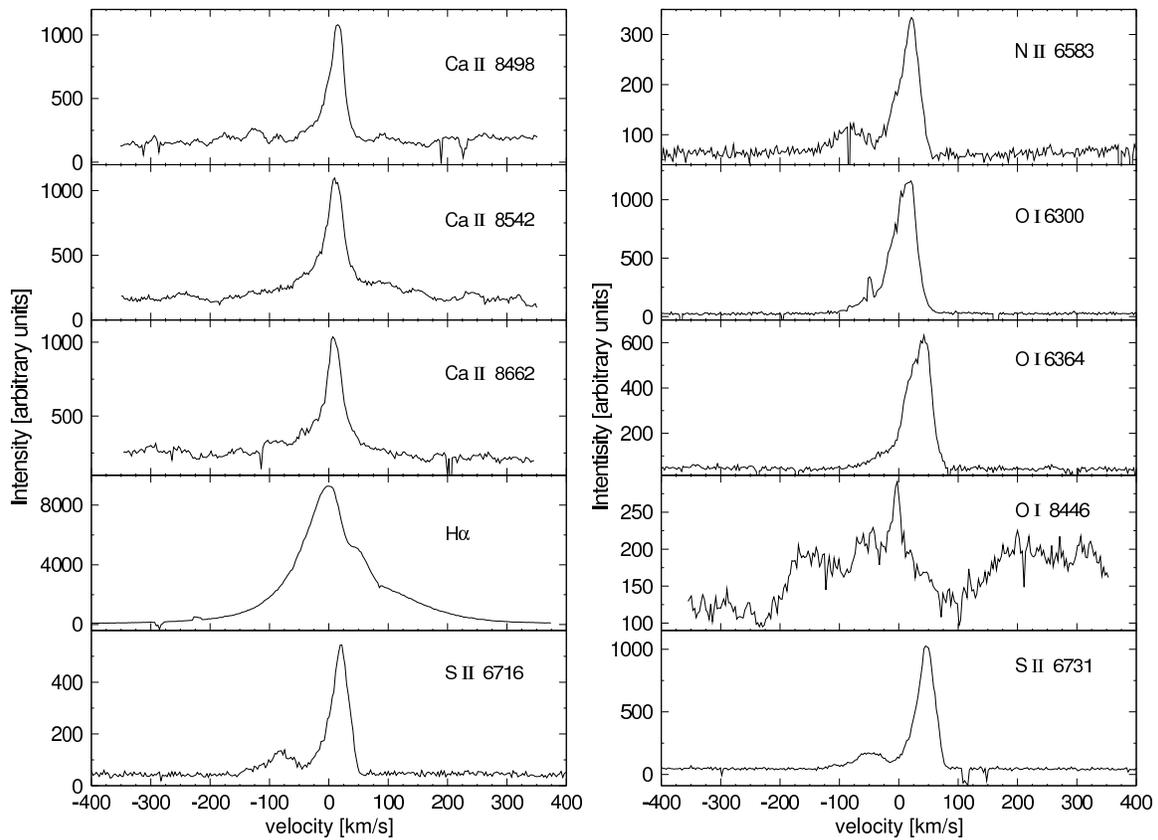}
      \caption{Emission lines in the Keck/HIRES spectra. Most of the lines are asymetric.}
         \label{lines}
   \end{figure*}

\noindent \emph{Accretion and mass loss. --} Both the low resolution CAHA/TWIN and high resolution Keck/HIRES spectra show strong emission lines superimposed on the stellar continuum (Fig. \ref{lines} and \ref{twin} and Table \ref{lines_table}). These lines are characteristic of intense accretion process: the H$_{\alpha }$ and \ion{Ca}{ii} triplet lines show equivalent widths typical of very active classical TTauri stars. A strong level of H$_{\alpha}$ emission had already been reported by \citet{2004ApJ...617.1216L} but no value of equivalent width was reported in their paper. The H$_{\alpha}$ line has a broad and asymmetric profile, as shown in Fig. \ref{lines}. We tentatively derive a mass accretion rate using the method described in \citet{2005ApJ...626..498M} based on the pseudo-equivalent width of the \ion{Ca}{ii} emission at $\lambda$8662~\AA, and using the spectral type and T$_{\rm eff}$ reported in \citet{2004ApJ...617.1216L} (M7.25, T$_{\rm eff}$=2838~K) and the corresponding mass for an age of 5~Myr as given by the DUSTY models \citep[M=0.045~M$_{\sun}$)][]{2002A&A...382..563B}. We obtain a value of $\log{\dot{\rm M}}$=-9.7~M$_{\sun}$ yr$^{-1}$, of the same order than observed for very low mass stellar and sub-stellar objects in Taurus \citep{2003ApJ...592..266M, 2003ApJ...582.1109W} and other associations of comparable ages \citep[e.g][]{2004ApJ...604..284B,2005ApJ...626..498M}. We note that the \ion{Ca}{ii} emission lines are observed in both HH objects and strongly accreting stars, so that the flux of this line might be common to both accretion and mass-loss and the mass accretion rate derived corresponds to an upper limit \citep{2005A&A...440..595A}. At the same time, numerous, strong forbidden optical lines (\ion{O}{i}, \ion{N}{ii} and \ion{S}{ii}) suggest that the object is also undergoing mass loss.

\section{Disk SED modelling}

We compare the SED of the target with a grid of 45\,000 passive disk models computed with a Monte Carlo Radiative Transfer code (MCFOST), fully described in \citet{2006A&A...459..797P}. In short, the code computes the disk temperature structure by propagating photon packets originally emitted by the central star through a combination of scattering following Mie theory, absorption and re-emission until they exit the computation grid and reach the observer. After the temperature computation, SEDs of the disk are estimated by emitting and propagating the proper amount of stellar and disk thermal photon packets, with a weight ensuring that each wavelength bin is properly converged. Ensuring energy conservation is a key issue for radiative transfer codes because of the very large surface densities reached in sub-stellar disk inner regions. Our two-steps calculations output two SEDs calculated by generating packets in two independent ways. The accuracy has been systematically a posteriori validated by comparing these two SEDs \citep[see][ section 3.3 for more details]{2006A&A...459..797P}. Synthetic SEDs are computed simultaneously at all inclinations in 10 bins regularly spaced in $\cos i$, with 60 wavelength bins logarithmically spaced between 0.1 and 4\,000~$\mu$m. The temperature computation is performed using 6.4$\times10^5$ packets, whereas synthetic SEDs are computed using approximately 6.4$\times10^6$ photons.

The parameter space for the disk model is strongly degenerated making it impracticable and too much time-consuming to explore a grid of sufficient resolution over all parameters. In order to reduce the size of the parameter space to be explored, we run preliminary models to get constraints on some of the parameters. The general shape of the silicates features is roughly reproduced  by amorphous olivine dust \citep{1995A&A...300..503D}. In the following global SED modeling approach, we do not try to get a perfect match of these features and adopt a composition of 100\% amorphous olivine. A detailed analysis of the mineralogy is performed afterwards.

As input we use the stellar parameters (T$_{\mathrm{eff}}$=2838~K, A$_{\mathrm{V}}$=0.0~mag, L=0.028~L$_{\sun}$) derived by \citet{2004ApJ...617.1216L}, the corresponding radius (R=0.64~R$_{\sun}$), the photospheric emission predicted by the NextGen models\footnote{for T$_{\rm eff}$=2\,800~K} of \citet{2002A&A...382..563B}, and an assumed outer disk radius of 300~AU. We assume a grain size distribution following a power-law with $dn(a)\propto a^{-3.5} da$,  between $a_{\rm min}$=0.03\,$\mu$m (chosen very small so that its exact value has no effect on the observables) and the maximum grain size  $a_{\rm max}$ considered as a free parameter. The spectral index between 1.3 and 3.47~mm $\log(F_\nu) /  \log(\nu) $ = 2.7 indicates the presence of grains significantly larger than interstellar ones. A maximum grain size around 1\,mm gives a good fit of the spectral index and we adopt $a_{\rm max} = 1\,$mm.  With this dust population, the millimeter emission is well reproduced with a dust mass of $10^{-5}$~M$_\odot$. Assuming a typical dust to mass ratio of 100, this leads to a total disk mass of $10^{-3}$\,M$_\odot$ in good agreement with the previously published estimate of 2.15$\times 10^{-3}$~M$_{\sun}$ \citep{2006ApJ...645.1498S}. 

The star properties, disk outer radius, grain size distribution, dust composition and dust disk mass are kept fixed in the rest of the modelling.  We then run a grid of models that systematically samples the remaining parameters: disk inclination, inner radius, scale height, flaring and surface density exponents. The explored ranges for each parameter are described in Table~\ref{parameters}.

\begin{table}
\caption{Explored parameters in our grid of models and results of the Bayesian analysis. \label{parameters}}
\begin{tabular}{lccccc}
\hline
Parameter & Min. value & Max. value & N$_{\rm sample}$ & sampling &
Valid range \\
\hline
R$_\mathrm{out}$ [AU] & 300  & 300  & 1 & fixed & \nodata \\
M$_\mathrm{dust}$ [M$_{\sun}$] & $10^{-5}$ & $10^{-5}$ & 1 & fixed & \nodata \\
a$_\mathrm{max}$ [mm] & 1 & 1 & 1 & fixed & \nodata \\
inclination [\degr] & 0 & 90 & 10 & linear in $\cos i$ & $\lesssim 55^\circ$ \\
surface density exp. & -2 & 0 & 5 & linear & -2 to -0.5\\
flaring index & 1.0 & 1.25 & 6 & linear & 1.1--1.2\\
R$_\textrm{in}$ [AU] & 0.015 & 0.15 & 6 & log & 0.02--0.1\\
H$_{0}$ [AU] & 20 & 80 & 13 & linear & 30--60\\
stratification & 0.& 0.1 & 5 & linear & no constraint\\
\hline
\end{tabular}
\end{table}

\begin{figure*}
   \centering
   \includegraphics[width=0.65\textwidth]{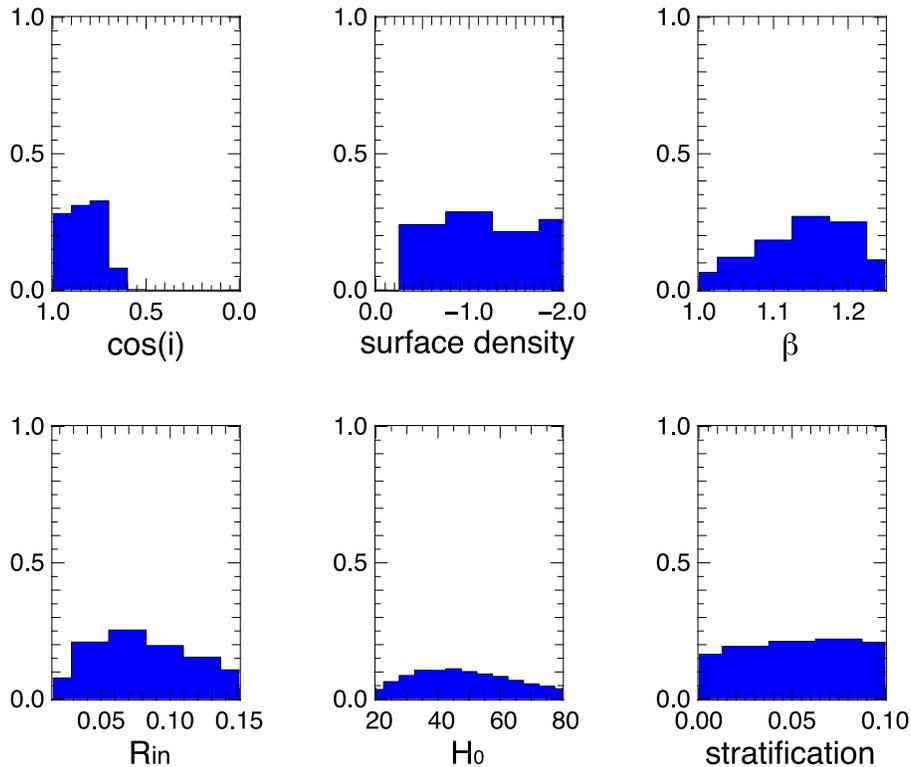}
      \caption{Distribution of probabilities of the output parameters of the \emph{MCFOST} fit. The inclination, $\beta$, and R$_{\rm in}$ have the best constraints, the other parameters being only loosely or not constrained. \label{proba}  }
   \end{figure*}

We performed a comparison of the synthetic SEDs with observations using a $\chi^2$ minimization. The best model is presented in Fig.~\ref{SED}. It gives an overall good match of all observational points from optical to millimeter regimes. As already mentioned, we did not intent to get a quantitative fit of the silicates emission bands but rather focus on the strength of these features. A Bayesian analysis is performed using the reduced $\chi^2$ to get an estimate of the validity range for each of the parameter explored  \citep{1992nrca.book.....P, 1997ApJ...489..917L, 2007A&A...469..963P}. A relative probability  $\exp (-\chi^2/2)$ is calculated for each individual model.  The relative likelihood for each of the parameters is obtained  by  adding the individual probabilities of all the models with a given parameter value.  These probabilities are the result of a marginalization of the parameter space successively over all dimensions and not a cut through the parameter space. In that sense, they account for potential correlations and interplay between parameters, and quantitative error bars can be extracted from them. The results are presented in Fig.~\ref{proba}.

\section{Disk mineralogy \label{mineralogy}}

The features present in the Spitzer IRS spectrum probe emission originating essentially from silicate dust grains less than a few microns in size (larger grains are featureless) and located in the optically thin surface layers of the inner disk ($<$2~AU). The following section presents a detailed analysis of the properties of these grains.

The data show several evidences of grain growth and processing. The slope of the SED at longer wavelengths indicates that large grains ($\approx$1~mm) must be present in the cold disk, while the trapezoidal silicate emission profiles seen in the Spitzer IRS spectrum around 10 and 18~$\mu$m are typical of grain growth to micron size and/or presence of crystalline silicate forsterite, the Mg-rich end member of the olivine group \citep{2001A&A...375..950B, 2005A&A...437..189V}. The IRS spectrum shows clear emission features at around 11.3, 23-24,  27-28, 33 microns that we attribute to crystalline silicate grains,  in particular crystalline forsterite which shows features at 10.1, 11.3, 16.2, 19.7, 23.8, 27.6 and 33.6~$\mu$m, and crystalline enstatite (Mg-rich pyroxene) which shows  features at 9.3, 10.6, 11.7, 19.6, 21.6, 23.0, 24.5 and 28.2~$\mu$m. The presence of crystalline grains in  emission indicates advanced processing of the circumstellar material  in the disk compared to the diffuse interstellar medium \citep{2004ApJ...609..826K}. In order to investigate the composition, distribution and level of processing of the grains of the inner disk (inner meaning here and after $\la$2~AU) of the target, we have followed the procedure used in interpreting the Spitzer IRS spectrum of a very low mass star by \citet{2007ApJ...661..361M}. Details of the procedure and references can be found therein. Briefly, the method includes three major steps:

\begin{enumerate}
\item subtraction of a continuum
\item fit of the continuum-subtracted 10~$\mu$m amorphous silicate feature (between 7--15~$\mu$m) using a grid of pre-calculated absorption efficiencies for five types of silicate species assumed to emit at the same T$_{\rm eff}$. A $\chi^{2}$ minimization provides the relative silicate mass fractions and best blackbody T$_{\rm eff}^{hot}$. 
\item fit of the residual spectrum (between 15 and 35$\mu$m) as in 2., providing a second, cooler blackbody T$_{\rm{eff}}^{cold}$ and relative mass fractions for the cold component.
\end{enumerate}

Following \citet{2007ApJ...661..361M} and references therein, the models includes five major grain species identified in circumstellar disks (3 types of amorphous silicates: olivine, pyroxene and silica; and 2 kinds of crystalline silicates: forsterite and enstatite), and assumes two representative (from a spectroscopic point of view) grain sizes (0.1 and 1.5~$\mu$m). The contribution of larger grains cannot be studied using the current dataset and would have been fitted in the continuum.

As mentioned in \citet{2007ApJ...661..361M}, the main limitation of this analysis lies in the degeneracy of the contributions of the grain species. The silicate emission indeed contributes to the emission over almost the entire IRS spectrum, preventing to measure accurately the purely photospheric+disk continuum emission which we need to subtract in step 1. To assess the uncertainties and limitations of this analysis, we investigate the mineralogy of the disk using independent methods for the continuum normalization.

\subsection{Continuum normalization}
 
\begin{figure*}
   \centering
   \includegraphics[width=\textwidth]{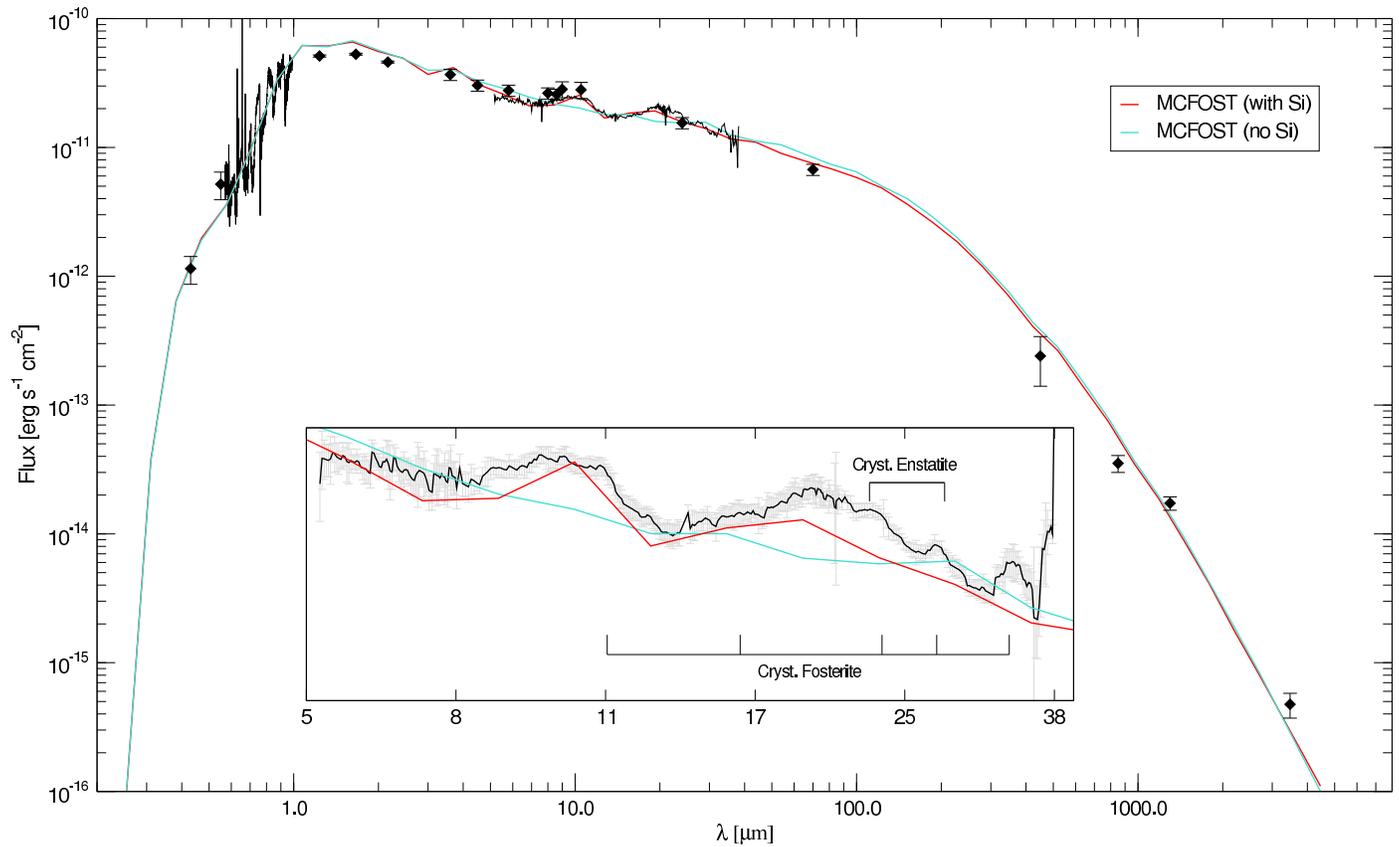}
      \caption{Spectral Energy Distribution of the target (black) and best fit \emph{MCFOST} model including the treatment of silicates (blue curve) and without silicates (red curve). The inset shows a zoom on the IRS spectrum. Some crystalline silicate features are indicated. \label{SED}  }
   \end{figure*}

\begin{figure*}
   \centering
   \includegraphics[width=0.95\textwidth]{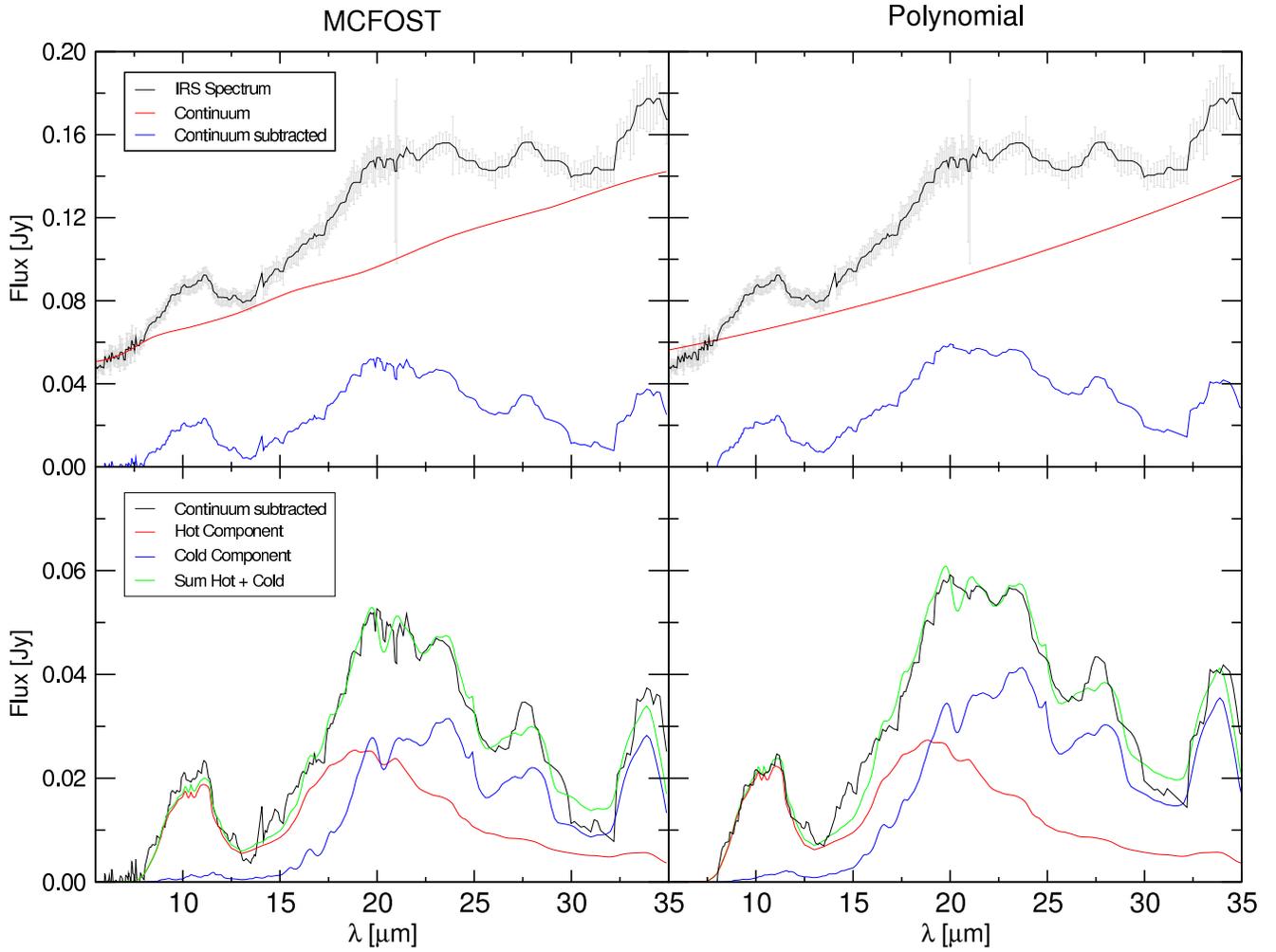}
      \caption{Upper Panels: Spitzer IRS spectrum (black), and the assumed continuum estimated using the best \emph{MCFOST} fit (left panel) and a polynomial fit (right panel) over-plotted in red, and the continuum subtracted flux (blue). Lower panels: continuum subtracted flux (black) and compositional analysis of the silicate emission. The blue line corresponds to the high temperature component, the red line to the low temperature component, and the green line to the sum of the two. \label{crystallinity}  }
   \end{figure*}

\noindent \emph{MCFOST disk model continuum. --} Our first approach consists in using the best-fit SED obtained with the \emph{MCFOST} code as the continuum. The main caveat is that a treatment of the optical properties of amorphous olivine is included in the \emph{MCFOST} models (see Fig. \ref{SED}, the 10 and 20~$\mu$m amorphous silicate features are relatively well fitted by the model), preventing to use the disk model as it is. In order to make a more reliable estimate of the continuum, we have computed the \emph{MCFOST} models without any treatment of the amorphous olivine emission by replacing the optical constants of this species with a power-law interpolation between 5 and 30~$\mu$m. The 10 and 20~$\mu$m silicate features are removed, but the energy originally re-emitted in the silicate features is now spread over the spectrum (the disk absorption indeed remains the same) and affects the continuum, as shown in Fig. \ref{SED}. In order to compensate this effect, we estimate the continuum using a smoothed version of this synthetic SED, shifted to match the 10~$\mu$m feature. The corresponding mineralogical analysis gives an overall good fit of the different features (see Tables \ref{table_cryst1} and \ref{table_cryst2}). The best fit is obtained for a mass fraction in crystalline grains of 25\%. Fig. \ref{crystallinity} gives an overview of the results.

\vspace{0.5cm}

\noindent \emph{Polynomial fit. --} We also derive an estimate of the continuum using a Lagrange polynomial fit following the method described in \citet{2006ApJ...639..275K} (in their Section 4.1, case 1). The fit obtained using this continuum gives a mass fraction in crystalline grains of 31\%, consistent with the previous estimate. The corresponding mineralogical analysis gives an overall good fit of the different features as shown in Fig. \ref{crystallinity} (right panels) and by the reduced-$\chi^{2}$ reported in Tables \ref{table_cryst1} and \ref{table_cryst2}.  As an additional test, we computed a second polynomial continuum with slightly different properties, leaving more flux around 35~$\mu$m. All the features, even the one at 33~$\mu$m, are well fitted assuming this second continuum. The estimated cristallinity, $\approx$20\%, is significantly lower than that estimated with the first polynomial continuum. Leaving more flux at longer wavelength naturally favors amorphous silicates, leading to a lower crystallinity fraction. The spectral features associated with amorphous silicates are indeed broader than that associated with crystalline silicates, and are therefore better suited to fill the flux deficit at long wavelengths in this continuum.

While the first polynomial fit of the continuum gives overall results very similar to those obtained with the \emph{MCFOST} models (see Fig. \ref{crystallinity}), the second polynomial fit gives a similarly good fit but with very different results, illustrating the difficulty of choosing the right continuum. Fortunately, the synthetic SED obtained with \emph{MCFOST} provides a first guess of the shape of the continuum, or at least of plausible range of acceptable continuum. The second polynomial fit is significantly different from the synthetic SED obtained with \emph{MCFOST}, and therefore less likely to be representative of the true continuum. It nevertheless provides an interesting check and a lower limit on the crystallinity level, at about 20\%. This iterative analysis, which takes advantage of both the \emph{MCFOST} and polynomial fits, leads us to conclude that the crystallinity mass fraction cannot reasonably be larger than $\approx$35\% or lower than $\approx$20\%, with the best fits obtained in the range between 25--31\%. The results of the analysis are summarized in Tables \ref{table_cryst1} and \ref{table_cryst2}.

\subsection{Grain sizes in the surface layers of the inner disk ($\lesssim$2~AU)}
Interestingly, the best fit is obtained for a mixture made of essentially intermediate size grains. We investigate the effect of the grain sizes on the results of the fit. The method originally uses two grain sizes: small grains $a_{\rm small}$=0.1~$\mu$m, and intermediate size grains $a_{\rm big}$=1.5~$\mu$m. Increasing the maximum grain size to 3.0~$\mu$m  gives a better fit of both the blue side of the 10 and 20~$\mu$m features, but a worse fit of their red sides, of the 29.2 and 30.9~$\mu$m features and of the relatively strong crystalline forsterite 33~$\mu$m feature. Larger grains ($\ge$6~$\mu$m) did not lead to any acceptable fit. We tentatively used 3 grain sizes, but the number of free parameters becomes too large and the results inconclusive. In conclusion, both the high and low temperature components are best fitted with $a_{\rm big}=$1.5~$\mu$m grains, consistent with the results of \citet{2006ApJ...639..275K} for more massive TTauri M stars. Smaller a$_{\rm big}$ produces narrower features, and larger a$_{\rm big}$ a poor fit of the red side of the different features. For the rest of the mineralogical analysis, we adopt a maximum silicate grain size of $a_{\rm big}$=1.5~$\mu$m. Finally, we stress that the 27.0~$\mu$m feature was never well fitted by any of the models mentioned above, must be associated to species not included in our analysis.

\section{Discussion}

\subsection{Disk structural properties}
The unique coverage and sampling of the SED allows to constrain a number of disk structural properties. 2MASS~J04442713+2512164 disk is the most massive disk known to date around a sub-stellar object, surpassing other observed brown dwarf disks by a factor 4 or more \citep{2006ApJ...645.1498S}.  This recalls the results observed for more massive TTauri stars showing that circumstellar disks of objects with well developed outflows are significantly more massive than disks of sources with no such outflows \citep{1990ApJ...354..687C}. Although no outflows was resolved in our NACO images, its presence is suggested by the numerous strong forbidden lines present in the optical spectrum, as mentioned above.

Fig. \ref{proba} presents the relative figure of merit estimated from the Bayesian inference of each disk parameter probed by our SED analysis with the MCFOST models. We obtain quantitative constraints on the inner radius, scale height and flaring exponent of the disk.  The inner radius is constrained between 0.02 and 0.1~AU with a peak probability around 0.04~AU. The flaring index is constrained between 1.1 and 1.2 and flat models $\beta=1$ are excluded. Acceptable values for the scale height are obtained in the range 30-60~AU, with a most probable value close to 45~AU. This large value for the scale height is consistent with the hydrostatic scale height expected for a brown dwarf \citep{2004MNRAS.351..607W}. The best model gives indeed a mid-plane temperature of 12.5~K at 100~AU. Assuming a central star mass of 0.045~M$_{\sun}$ and a vertically isothermal disk (which is not the case in the model), the mid-plane temperature translates in a scale height $\sqrt{k_B r^3 T(r) / G M_{\star} \mu} \approx 35$~AU. The increasing temperature towards the disk surface will result in slightly a larger "effective" scale height, therefore in good agreement with the scale height deduced from the SED fitting. All these properties are very similar to the properties of disks of more massive TTauri, suggesting that similar processes are at work on each side of the sub-stellar limit.

Some of the explored parameters cannot be constrained by our modelling of the SED, resulting in flat probabilities. The settling level is almost unconstrained and good fits can be found with and without settling. This is consistent with the fact that the dust scale height is in agreement with the gas scale height estimated assuming hydrostatic equilibrium and indicates that the dust and gas are probably well mixed within the disk. The surface density exponent is also only loosely constrained. Models with flat surface densities are excluded, the opacity in the central regions of the disk becoming too low to produce enough near-infrared excess. The fit is also insensitive to the disk inclination as long as the star remains directly visible by the observer. When the inclination becomes too large ($\gtrsim 55^\circ$), the star starts to be occulted by the disk and the optical and near-infrared fluxes are strongly reduced. 

\begin{figure*}
   \centering
   \includegraphics[width=0.5\textwidth]{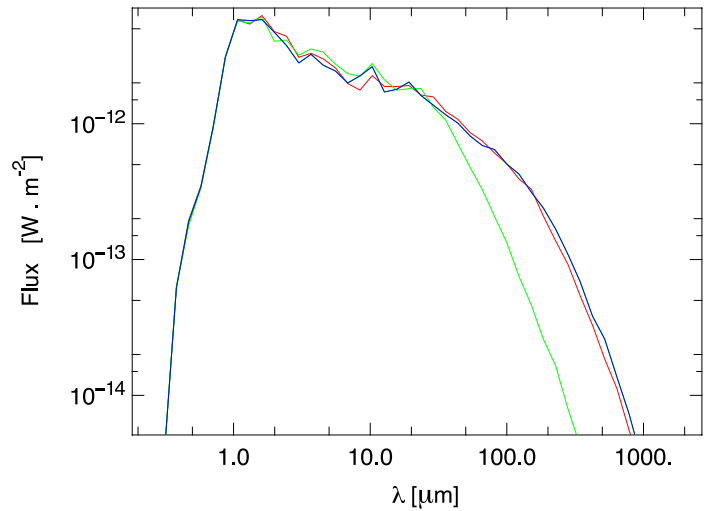}
      \caption{Synthetic SED computed with \emph{MCFOST} for R$_{\mathrm{out}}$=300, 10, 1~AU (blue, red, and green, respectively), all the other parameters being free. Fits of similar goodness can be obtained for any values of R$_{\mathrm{out}}>$10~AU. When the outer radius becomes too small, the disk becomes optically thick and the flux in the sub-mm/mm range decreases dramatically. \label{SED_rout}  }
   \end{figure*}

We also tentatively investigate the external disk truncation via the influence of the outer radius on the fit by computing synthetic SEDs for three different values of R$_{\mathrm{out}}$. Fig. \ref{SED_rout} shows synthetic SEDs obtained for R$_{\mathrm{out}}=$300, 10 and 1~AU. The SED remains almost unchanged down to an outer radius as small as 10~AU, while keeping the disk mass constant. At R$_{\mathrm{out}}=$1~AU, the flux in the sub-mm/mm range decreases dramatically, the disk becoming optically thick, and indicating that in this case the limit must lie between 1 and 10~AU. This emphasizes a limitation of such analysis and the fact that the outer radius and disk truncation cannot be directly constrained from the sub-millimeter/millimeter photometry. A way to break this degeneracy is to image the disk, as illustrated recently in \citet{2007ApJ...666.1219L} for another sub-stellar member of the Taurus association. We obtained deep adaptive optics images with the aim to search for extended diffuse emission coming from the disk or the jet. The faintness of the source, and the intermediate inclination of its disk made these observations and the data analysis challenging. The current results are negative and no obvious extension is found in the images.

\subsection{Crystallization of Silicates}

Table \ref{table_cryst1} gives a summary of the main properties of the 0.1$<a<$1.5~$\mu$m population of grains in the hot and cold temperature components of the optically thin layers of the inner disk ($\la$2~AU) probed by the mid-IR Spitzer spectrum. Both components seem dominated by intermediate size grains ($a$=1.5~$\mu$m), indicating a relatively high level of processing. This is consistent with the results of \citet{2006ApJ...639..275K} who report a statistical dependency of grain size on spectral type for more massive TTauri stars, cooler objects harboring disks made of larger grains.  

The hot and cold temperature components probed by the IRS spectrum are produced in different locations of the inner disk, the cold component corresponding mostly to the outer part and/or deeper layers of the inner disk, and the hot component corresponding to the surface layers and/or inner region of the inner disk (see Fig. \ref{region_emission}). Interestingly, the cold temperature component contains a significantly higher fraction of crystalline species than the hot component (10 to 15 times larger).  Since the cold component contributes to about half of the mass, as shown in Table \ref{table_cryst2}, it implies that crystalline species are mostly located in the deeper layers/outer region of the disk. This result does not fit in the classical picture where thermal annealing of amorphous silicates takes place by means of heating to high temperatures very close to the central star \citep[at about 800~K, corresponding to 10$^{-1}\sim$10$^{-3}$~AU in the case of brown dwarfs, ][]{1998A&A...332.1099G, 2000ApJ...535..247H,2007ApJ...659..680K} and resulting in a higher crystallinity closer to the star than in the outer parts of the disks \citep{2004Natur.432..479V}. In the following sections we discuss our results in the context of tree scenarios that may be responsible for the production of crystalline silicate grains in protoplanetary disks: (1) annealing in the very early stages of the accretion disk formation; (2) annealing in the hot inner regions of the disk followed by large-scale outward radial transport; (3) local annealing in the outer cold regions produced by energetic events such as flares or shocks. 

\subsubsection{Thermal annealing in the early phase of disk formation}
\citet{2006ApJ...640L..67D} suggest that dust crystallization events happen in the very early stages of the disk formation and relate the abundance of crystalline species in the entire disk to the rotational velocity of the parent cloud. Rapidly rotating clouds produce massive disks with lower accretion rates and lower crystallinity, while slow rotating clouds produce less massive disks with lower accretion rate but high levels of crystallinity. This process would fill the protoplanetary disk with a relatively homogeneous radial distribution of crystalline species. Fig. \ref{crystfracacrre} shows several cases that do not fit in this scheme (larger accretion rates but significant levels of crystallinity), including 2MASS~J04442713+2512164 which additionally has the most massive disk known to date around a brown dwarf. If at work, this scenario must therefore be followed by additional annealing and/or mixing processes as the disk evolves with time, and the question arises as to what explains the larger fraction of crystalline species among the 0.1$<a<$1.5~$\mu$m population of grains in the deeper layers/outer regions of the inner disk probed by our study.  

\subsubsection{Thermal heating in the disk inner regions and radial mixing} 

\citet{2002A&A...384.1107B} and then \citet{2004A&A...415.1177K} described models were turbulent diffusion carries crystalline silicates from the inner to the outer disk, while the surface layers are slowly spiraling inwards to reach high temperatures and anneal. The accretion rates required in their simulations to transport the annealed grains to the outer regions were inconsistent with the most recent models of formation of the solar system. Slower accretion rates were not able to reproduce the observed abundances of processed grains at large radii mostly because the outward transport of processed grains was surpassed by the inward flow of material accreting onto the central star. \citet{2007Sci...318..613C} recently developed further these studies and presented a two-dimensional protoplanetary disk model including the treatment of both the radial and the vertical transport of grains. The new simulations show that the outward transport of processed grains is a stratified mechanism which appears to be more efficient than initially predicted by the previous one-dimensional models. Outward transport occurs most efficiently near the disk mid-plane, where the pressure due to the gradient of density eases the transport to the outer regions, and where the grains do not have to fight the inward flow of accreting material. Accretion plays a key role in these scenarios. The efficiency and spatial scales on which these additional mechanisms are taking place are difficult to quantify with the current data, accretion being a strongly variable phenomenon, and the scales probed by mid-IR spectroscopy depending on the geometry of the disk (see discussion in section \ref{previousstudies}). However, recent statistical studies carried over a large sample of solar mass TTauri by \citet{2007ApJ...659..680K} showed no correlation between the grain properties and the accretion as measured by the H$_{\alpha}$ equivalent width. Fig. \ref{crystfracacrre} shows the crystalline mass fraction of samples of Herbig Ae/Be stars, TTauri stars and brown dwarfs found in the literature. Even though the limited size and the inhomogeneity of the samples (spanning different ages, star forming regions and measurement methods) represented in this figure prevent us to draw any firm conclusion, there is no indication of a correlation between the crystalline mass fraction and the accretion rate on either side of the sub-stellar limit.

Finally, if the mixing is not efficient enough, the lower crystalline mass fraction in the hot component of 2MASS~J04442713+2512164 could be explained by reverse transformation (melting) of crystalline to amorphous species if the newly formed crystals are not transported fast enough to deeper layers or outer radii.

\subsubsection{Transient grain annealing}
Energetic events (shocks in the disk or flaring of the central object) could be responsible for a significant production of crystalline silicates in outer regions where the temperature is much cooler than the annealing temperature. \citet{2002ApJ...565L.109H} computed the elevation of temperature due to shock waves triggered by gravitational instabilities in the case of typical TTauri systems and concluded that annealing of silicate grains in the 5--10~AU region is possible. Because of their smaller thermal resistance, small grains are more likely to be annealed and the fraction of crystalline species is therefore expected to be larger for small grains that for large grains. Our analysis of the mid-IR spectrum of 2MASS~J04442713+2512164 shows that the small grains of the hot components are indeed mostly crystalline, but that they are mostly amorphous in the cold component. Spectroscopy at longer wavelengths \citep[with e.g the Herschel Space Observatory,][]{2004AAS...204.8101P} should bring new insights on the contribution of transient annealing in protoplanetary disks.

\subsection{Additional crystallization processes}
\citet{1999Natur.401..563M} have discussed several mechanisms to increase the degree of crystallinity of red giant disks without the need for radial mixing: selective removal of amorphous grains (either by transport or coagulation) or additional crystallization methods (below the annealing temperature). Their conclusion is that the latter explanation is the most likely one, selective transport occurring on much slower timescales and coagulation affecting crystalline and amorphous species in the same manner. Because of the similarities between red giant and young star disks, they also conclude that such additional mechanism(s) could be at work in young star disks. Recent laboratory experiments have shown that exothermic chemical reactions involving the graphitization of a carboneous mantle layer covering amorphous silicate grains are able to transform these grains into crystalline silicate grains \citep{2007ApJ...666L..57K}. Both such organic species and crystalline silicate grains have been found in the comet 81P/Wild 2 samples returned by the Stardust mission \citep{2006Sci...314.1728K} and in the observations of comet 9P/Tempel 1 after the Deep Impact experiment \citep{2007Icar..191..432H,2007A&A...476..979T}.

\subsection{Grain growth}

\begin{figure*}
   \centering
   \includegraphics[width=0.65\textwidth]{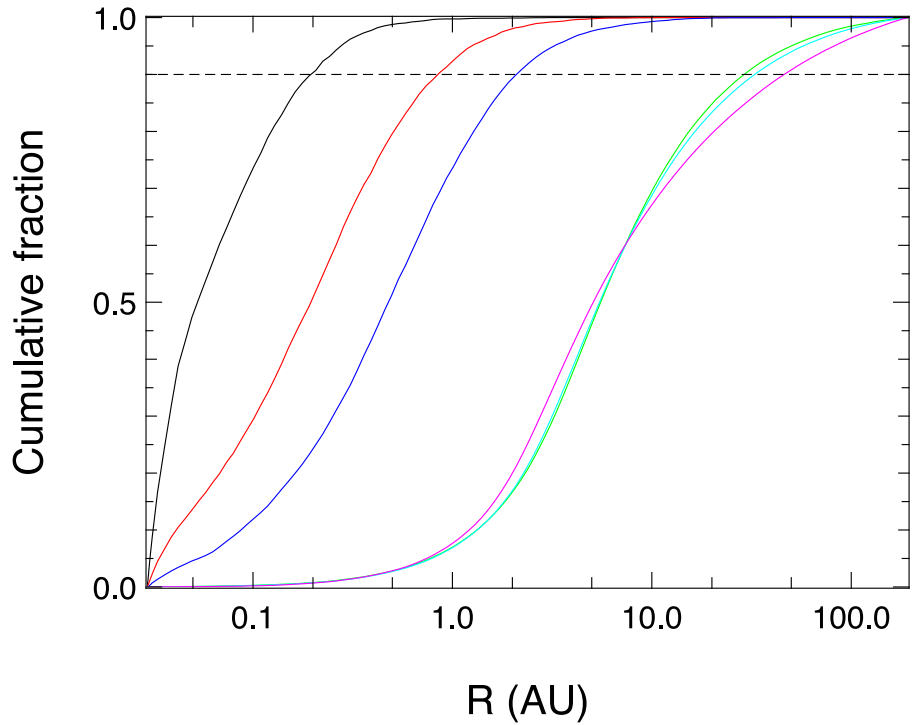}
    \caption{Cumulative flux at 10~$\mu$m (black), 20~$\mu$m (red), 35~$\mu$m (blue) 350~$\mu$m (green), 450~$\mu$m (cyan) and 1300~$\mu$m (magenta) as a function of the radius. Fluxes are calculated using the \emph{MCFOST} models of \citet{2006A&A...459..797P} for 2MASS~J04442713+2512164 best fit model. Of the different fluxes, 90\% (horizontal dashed line) comes from within $\approx$0.2~AU at 10~$\mu$m, 1$\sim$2~AU at 10 and 20~$\mu$m, and $>$20~AU for $>$350~$\mu$m. \label{region_emission}}
   \end{figure*}

Coagulation into large millimeter size grains is clearly demonstrated by the sub-millimeter and millimeter fluxes of 2MASS~J04442713+2512164. Millimeter photometry mostly traces the thermal emission of cold grains in the outer part ($>$20~AU) of the disk mid-plane, as shown in Fig. \ref{region_emission}. The presence of silicate features in the mid-IR spectrum shows that intermediate size grains ($a\approx$1~$\mu$m) are also still present in the inner disk ($\lesssim$2~AU, see Fig. \ref{region_emission}). On the other hand, the same mid-IR spectral analysis shows a lack of small grains ($a\approx$0.1~$\mu$m) in these regions/surface layers (see Table \ref{table_cryst1}). These three results together with the much higher level of crystallinity found in the cold component of the inner disk show that the disk of 2MASS~J04442713+2512164 has reached a relatively advanced level of processing, leading to the formation of elaborate crystalline structures and large grains. 

Unfortunately, the non-detection of the disk in direct imaging and the degeneracy of the SED modelling of the optically thin disk at sub-mm and mm wavelengths do not allow to derive any constraints on the vertical and radial distribution of the large grains. Fig. \ref{proba} shows that in the current state of the data and of the models, no constraints can be derived on the vertical distribution of the grains. Simple considerations on the dynamics of large dust grains, which is mainly determined by gravitational sedimentation towards the disk mid-plane, naturally suggest a stratified disk structure with large grains  mostly located in the disk mid-plane, and smaller grains at the surface of the disk \citep{2001ApJ...547.1077C}. This simple picture can nevertheless be complicated by radial and vertical mixing, and fragmentation of the grains via collisions. The larger crystalline mass fraction of the cold component suggest that such settling is at work and must be very efficient. Direct imaging of the disk at different wavelengths and/or interferometric observations are required to constrain the grain size radial and vertical distributions.

\subsection{Comparison with previous studies and limitations of the analysis \label{previousstudies}}

The methods used to model the dust properties suffer from several limitations. None of the models in this paper (or in the literature) provide a self-consistent and physically correct dust model, mainly due to several unconstrained parameters or to simplifications. These include the limited number of temperatures (two), the limited number of grain species (both in terms of sizes and nature) and in particular the important effects of dust species without infrared features on the continuum, cold grains, and the optically thin continuum from large ($>$10~$\mu$m) silicate grains that do not show up as spectral features, and the porosity of the grains. The values derived in this paper should therefore be taken with great caution. Keeping these limitations in mind, our analysis nevertheless allows to perform qualitative comparisons with other objects observed and studied using similar methodology.

\begin{figure*}
   \centering
   \includegraphics[width=0.6\textwidth]{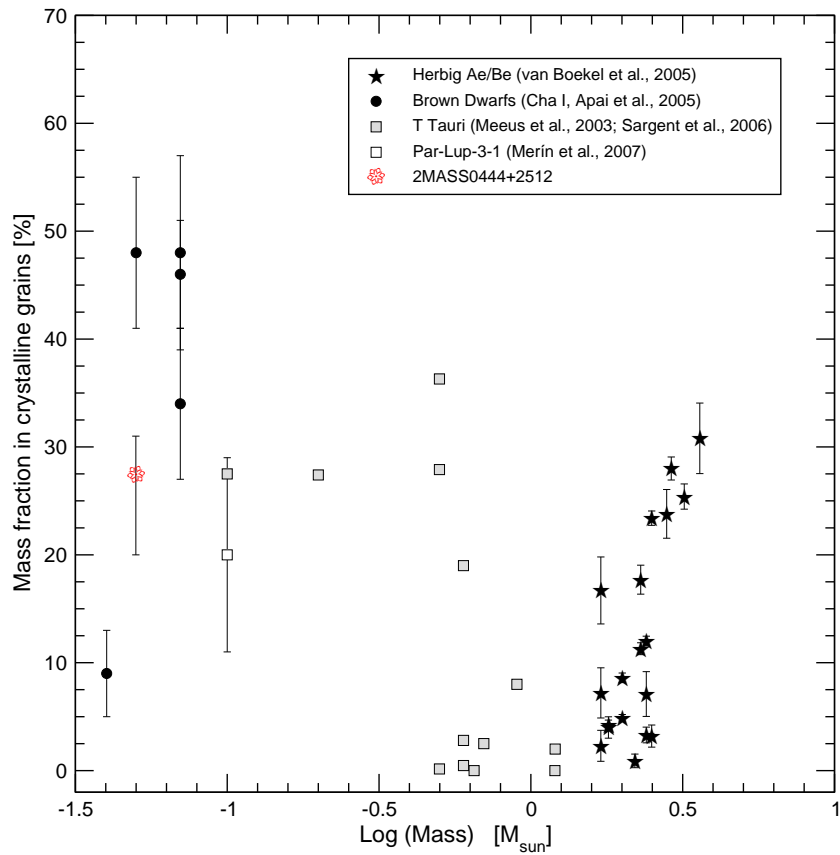}
      \caption{Grain evolution on stellar parameter: Crystalline mass fraction of the inner disk ($\lesssim$2~AU) as a function of the mass for stellar and substellar objects \citep[Fig. from ][ updated with new measurements]{2005Sci...310..834A}. 2MASS~J04442713+2512164 is represented with a red flower. \label{massfraccryst}  }
   \end{figure*}

\begin{figure*}
   \centering
   \includegraphics[width=0.6\textwidth]{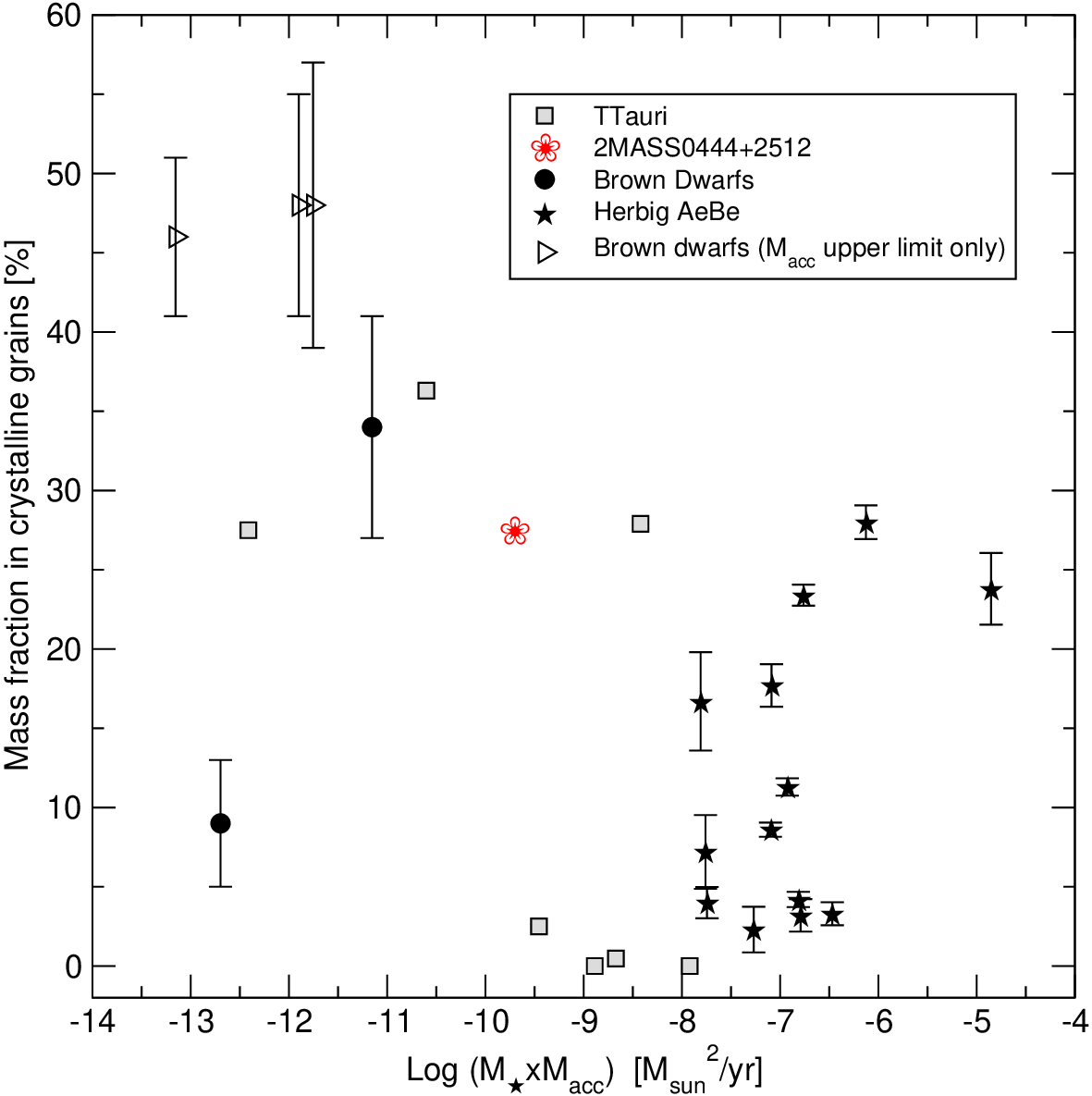}
      \caption{Grain evolution on disk parameter: crystalline mass fraction of the inner disk ($\lesssim$2~AU) as a function of the accretion luminosity (L$_{\rm acc} \propto$ M$_{\star}\times\dot{\rm M}$) for Herbig Ae/Be stars \citep[stars][]{2005A&A...437..189V} and brown dwarfs \citep[][filled circle if a mass accretion rate has been measured, and open right-triangle if only a limit is available]{2005Sci...310..834A} and TTauri stars \citep[][]{2006ApJ...645..395S} for the objects represented in Fig. \ref{massfraccryst}. 2MASS~J04442713+2512164 is represented with a red flower. In the current state of the observations there seem to be no correlation between the mass accretion rate and the crystallinity on either side of the substellar limit. Mass accretion rate measurements from \citet{2006A&A...459..837G,2005ApJ...626..498M,2005ApJ...625..906M,2000ApJ...545L.141M,2007AAS...211.1220G,2005ApJ...630L.185C} and this work. \label{crystfracacrre}  }
   \end{figure*}

Fig. \ref{massfraccryst} compares the mass fraction in crystalline grains of 2MASS~J04442713+2512164  to that of more massive Herbig AeBe stars \citep{2005A&A...437..189V}, TTauri stars \citep{2003A&A...409L..25M,2006ApJ...645..395S,2007ApJ...661..361M} and other brown dwarfs \citep{2005Sci...310..834A}. With a crystalline mass fraction between 20--30\% and a $S_{11.3}/S_{9.3}$ ratio equal to 1.05, 2MASS~J04442713+2512164 has an overall degree of crystallization comparable to objects of similar mass and ages. 

The compositional analysis of the M5.5 star Par-Lup-3-1 performed by \citet{2007ApJ...661..361M} provides an opportunity to compare the results obtained for two objects close to the sub-stellar boundary and with similar ages (the estimated ages of the Lupus~III and Taurus associations being around 1--3~Myr). \citet{2007ApJ...661..361M} report a much larger fraction of small grains (0.1~$\mu$m) in both the hot and cold components, contrasting with the lack of such small grains found in our analysis of 2MASS~J04442713+2512164 spectrum. The millimeter observations being not available for Par-Lup-3-1, the following comparison between the properties of the two disks holds for the inner disk ($\lesssim$2~AU) probed by the mid-IR analysis. The hot component of Par-Lup-3-1 contains most of the crystalline mass, while we find a much higher crystalline mass fraction in the cold component of 2MASS~J04442713+2512164. The most striking difference between these two objects lies in the lack of any accretion signature in the optical spectrum of Par-Lup-3-1 \citep{2003A&A...406.1001C}, while 2MASS~J04442713+2512164 displays one of the strongest accretion related H$\alpha$ luminosities. Fig. \ref{crystfracacrre} shows nevertheless that there seem to be no correlation between the accretion luminosity and the crystallinity of sub-stellar and very low mass objects. Several other independent parameters could well be responsible for the differences observed in 2MASS~J04442713+2512164  and Par-Lup-3-1 disk compositions. The initial environmental conditions or the disk processing mechanisms at work on each side of the sub-stellar limit must indeed play crucial roles as well \citep{2006ApJ...640L..67D} and are not taken into account in the current analysis. The disk geometries, and in particular the orientation, must also affect the results of our analysis. The smaller inclination of 2MASS~J04442713+2512164 ($i<$55\degr\, instead of $i>$70\degr\, for Par-Lup-3-1) possibly implies that the mid-IR spectrum probes larger depths into the low-density surface layers, ``\emph{artificially}" increasing the overall mass fraction in the cold component when compared to Par-Lup-3-1. On the other hand, a smaller inclination also corresponds to a smaller projected surface of the internal edge, where most of the mid-IR emission probed by the mid-IR Spitzer spectrum originates. These opposing geometrical effects are not fully taken into account in the IRS spectral analysis, and not only complicate a direct comparison between different sources but also add an unknown weighting to the overall compositional analysis. Similar analysis on large samples covering a statistically well defined range of geometries are required to draw more quantitative conclusions and comparisons.

\section{Conclusions}

We have presented a complete analysis of a brown dwarf SED from the optical to the millimeter, and a detailed compositional analysis of its entire (5--35~$\mu$m) mid-IR Spitzer spectrum. The conclusions are severalfold:
\begin{enumerate}
\item The disk has geometrical properties similar to that observed for more massive TTauri. In the current state of the models and of the observations, the analysis of the SED does not allow to derive any constraints regarding the outer radius of brown dwarf disks. Millimeter observations lack the angular resolution to study disk truncation. Additional observations including direct imaging, and spectroscopy of the inner region via interferometry or e.g with the future Herschel Observatory, are required in order to study the vertical and radial distribution of grains, but also to search for evidence of disk truncation.
\item The sub-mm and mm photometry indicates that most of the mass of the disk is in grains significantly larger ($\lesssim$1~mm) than in the diffuse interstellar medium (100\AA--0.2~$\mu$m).  Grain growth to millimeter size grains can therefore occur as rapidly as 1--3~Myrs. 
\item As for the very low mass M5.5 star Par-Lup-3-1, the mid-IR Spitzer spectrum is better fitted using two temperature components rather than a single one. The hot component displays a significantly lower crystallinity than the cold component, indicating an advanced level of processing, and possibly contrasting with Par-Lup-3-1. This also suggests that: (i) the radial and vertical mixing of crystalline species in the disk could be very efficient; (ii) that crystallization processes other than high temperature annealing close to the central object might be at work; (iii) that crystals might be melted back to amorphous state faster than they are transported out of the hot layers where they form
\item In the current state of the observations, the composition of very low mass stellar and sub-stellar disks does not seem to correlate with accretion.
\item The proportions of intermediate size grains (1.5~$\mu$m) in both the hot and cold components are larger than those derive for Par-Lup-3-1, providing additional evidence of a more advanced level of processing than the M5.5 very low mass star even though the two objects are located in associations of similar ages (1--3~Myrs). 
\end{enumerate}

These results provide additional evidences that brown dwarfs disks are similar to those of more massive TTauri stars, and that the initial steps of planet formation are present during the process \citep{2005Sci...310..834A}.

\begin{acknowledgements}
H. Bouy acknowledges the funding from the European Commission's Sixth Framework Program as a Marie Curie Outgoing International Fellow (MOIF-CT-2005-8389). We are grateful to our referee Daniel Apai for his comments and advice which helped improve this manuscript considerably.

This work is based on observations obtained at the VLT which is operated by the European Southern Observatory. This work makes use of DENIS data. The DENIS project has been partly funded by the SCIENCE and the HCM plans of the European Commission under grants CT920791 and CT940627. It is supported by INSU, MEN and CNRS in France, by the State of Baden-W\"urttemberg in Germany, by DGICYT in Spain, by CNR in Italy, by FFwFBWF in Austria, by FAPESP in Brazil, by OTKA grants F-4239 and F-013990 in Hungary, and by the ESO C\&EE grant A-04-046. This publication makes use of data products from the Two Micron All Sky Survey, which is a joint project of the University of Massachusetts and the Infrared Processing and Analysis Center/California Institute of Technology, funded by the National Aeronautics and Space Administration and the National Science Foundation. This work is based [in part] on archival data obtained with the Spitzer Space Telescope, which is operated by the Jet Propulsion Laboratory, California Institute of Technology under a contract with NASA. The IRS was a collaborative venture between Cornell University and Ball Aerospace Corporation funded by NASA through the Jet Propulsion Laboratory and Ames Research Center. The authors acknowledge the data analysis facilities provided by the Starlink Project which was run by CCLRC on behalf of PPARC. This work has made use of the Vizier Service provided by the Centre de Donn\'ees Astronomiques de Strasbourg, France \citep{Vizier}. This research used the facilities of the Canadian Astronomy Data Centre operated by the National Research Council of Canada with the support of the Canadian Space Agency. Some of The data presented herein were obtained at the W.M. Keck Observatory, which is operated as a scientific partnership among the California Institute of Technology, the University of California and the National Aeronautics and Space Administration. The Observatory was made possible by the generous financial support of the W.M. Keck Foundation. The authors wish to recognize and acknowledge the very significant cultural role and reverence that the summit of Mauna Kea has always had within the indigenous Hawaiian community.  We are most fortunate to have the opportunity to conduct observations from this mountain. Some of The data presented herein were obtained at the Calar Alto Observatory, which is operated jointly by the Max-Planck-Institut f\"ur Astronomie (MPIA) in Heidelberg, Germany, and the Instituto de Astrof\'\i sica de Andaluc\'\i a (CSIC) in Granada/Spain.

\end{acknowledgements}

\begin{center}
\begin{deluxetable}{lcc }
\tablecaption{Photometry \label{photom}}
\tablewidth{0pt}
\tablehead{
\colhead{$\lambda$}      & \colhead{Flux}      & \colhead{Ref.} \\
\colhead{[$\mu$m]}        & \colhead{[mJy]}     & 
}
\startdata
0.43                     & 0.081$\pm$0.022     &       (1)        \\ 
0.55                     & 0.264$\pm$0.068     &       (1)        \\ 
0.7                      & 0.466$\pm$0.096     &       (1)        \\
1.235                    & 21.1$\pm$0.4        &       (2)        \\
1.662                    & 29.3$\pm$0.5        &       (2)        \\
2.159                    & 33.1$\pm$0.5        &       (2)        \\
3.6                      & 44.1$\pm$4.4        &       (3)        \\
4.5                      & 45.5$\pm$4.5        &       (3)        \\
5.8                      & 53.5$\pm$5.3        &       (3)        \\
8.0                      & 70.3$\pm$7.0        &       (3)        \\
8.59                     & 73.9$\pm$7.1        &       (5)        \\
8.99                     & 84.8$\pm$12.0       &       (5)        \\
10.49                    & 97.9$\pm$14.0       &       (5)        \\
24                       & 124.0$\pm$12.4      &       (3)        \\
70                       & 157.0$\pm$15.7      &       (3)        \\
450                      & 36$\pm$15           &       (5)        \\
850                      & 10$\pm$1.5          &       (5)        \\
1300                     & 7.5$\pm$0.89        &       (4)        \\
3470                     & 0.55$\pm$0.12       &       (5)        \\
\enddata
\tablerefs{(1) \citet{2005yCat.1297....0Z}; 
(2) 2MASS;  
(3) \citet{2007A&A...465..855G}; 
(4) \citet{2006ApJ...645.1498S}; 
(5) this work}
\end{deluxetable}
\end{center}

\begin{center}
\begin{deluxetable}{lccc }
\tablecaption{Pseudo-equivalent widths of emission lines (in \AA) \label{lines_table}}
\tablewidth{0pt}
\tablehead{
\colhead{Line}      & \colhead{Keck/HIRES}  &   \colhead{CAHA/TWIN} \\
                    & \colhead{(R=31\,000, 2006 Oct. 13)}  & \colhead{(R=3\,300, 2006 Nov. 11)} }
\startdata
\ion{O}{i}   6300 & 48.3$\pm$4.5  & 72$\pm$3 \\
\ion{O}{i}   6364 & 15.8$\pm$ 0.5 & 17$\pm$2 \\
H$\alpha$    6563 & 620$\pm$330\tablenotemark{1}   & 124$\pm$20   \\
\ion{N}{ii}  6581 & 5.2$\pm$1.4   & 8.7$\pm$2.0 \\
\ion{S}{ii}  6717 & 13.0$\pm$0.6  & 14$\pm$3 \\
\ion{S}{ii}  6731 & 22.5$\pm$1.4  & 30$\pm$2 \\
\ion{Ca}{ii} 8498 & 5.4$\pm$0.3   & \nodata \\
\ion{Ca}{ii} 8542 & 3.4$\pm$0.1   & \nodata \\
\ion{Ca}{ii} 8662 & 2.6$\pm$0.2   & \nodata \\
\enddata
\tablenotetext{1}{The equivalent width of such a strong, broad and asymetric line is only indicative}
\end{deluxetable}
\end{center}

\begin{deluxetable}{lcc }
\tablecaption{VLT/VISIR Filters properties \label{visir_table}}
\tablewidth{0pt}
\tablehead{
  \colhead{Filter} & \colhead{Central Wavelength}      & \colhead{Half-band width}\\
\colhead{}         & \colhead{[$\mu$m]}                & \colhead{[$\mu$m]}  
}
\startdata
PAH1 	& 8.59 	& 0.42 	\\
\ion{Ar}{iii} 	& 8.99 	& 0.14 	\\ 	
\ion{S}{iv} 	& 10.49 & 0.16 	\\ 	
PAH2 	& 11.25 & 0.59 	\\ 	
\ion{Ne}{ii} 	& 12.81 & 0.21 	\\ 	
Q1 	& 17.65 & 0.83 	\\ 	
Q2 	& 18.72 & 0.88 	\\
\enddata
\end{deluxetable}

\begin{deluxetable}{lc}
\centering
\tablecaption{Properties of 2MASS~J04442713+2512164 \label{table_properties}}
\tablewidth{0pt}
\tablehead{\multicolumn{2}{c}{Stellar properties}}
\startdata
SpT                    & M7.25$\pm$0.25 \\
$v \sin{i}$            & 12$\pm$2~km $\rm s^{-1}$ \\
veiling $r_{6750\AA}$  & 0.4$\pm$0.1 \\
A$_{\rm V}$            & 0.0~mag \\
$\log{\dot{\rm M}}$ [M$_{\sun}$/yr] & -9.7 \\
\enddata
\end{deluxetable}

\begin{deluxetable}{lccccc}
\centering
\tablecaption{Mineralogical composition of the inner disk probed by our analysis \label{table_cryst1}}
\tablewidth{0pt}
\tablehead{
  \colhead{Continuum} & \colhead{$a_{\rm small}=$0.1~$\mu$m}      & \colhead{$a_{\rm big}=$1.5~$\mu$m} & \colhead{Crystalline mass fraction} & \colhead{Temperature} & \colhead{$\chi^{2}$}\\
\colhead{}         & \colhead{[\%]}                & \colhead{[\%]} & \colhead{[\%]} & \colhead{[K]}  
}
\startdata
\multicolumn{6}{c}{High Temperature Component} \\
\hline
Polynomial   & 4.4  & 95.6   & 4.4  & 256 & 0.12 \\
MCFOST       & 5.5  & 94.5   & 5.5  & 267 & 0.19 \\
\hline
\multicolumn{6}{c}{Low Temperature Component} \\
\hline
Polynomial   & 10.1      & 89.9       & 42.3 & 127 & 0.35 \\
MCFOST       & 59.5      & 40.5       & 66.4 & 133 & 0.30 \\
\hline
\multicolumn{6}{c}{High + Low Components} \\
\hline
Polynomial   & 7.6     & 92.4       & 25.4 & \nodata  & 1.51 \\
MCFOST       & 30.3    & 69.7       & 33.5 & \nodata  & 0.31 \\
\enddata
\end{deluxetable}

\begin{deluxetable}{lccc}
\centering
\tablecaption{ Fractional masses of the two components\label{table_cryst2}}
\tablewidth{0pt}
\tablehead{
  \colhead{Continuum} & \colhead{Hot Comp.} & \colhead{Cold Comp.} \\
\colhead{}            & \colhead{[\%]}      & \colhead{[\%]}
}
\startdata
Polynomial  &  44.5         &       55.5  \\
MCFOST      &  54.0         &       46.0  \\
\enddata
\end{deluxetable}

\bibliographystyle{aa}

\end{document}